\documentclass[10pt,conference]{cls/IEEEtran}
\IEEEoverridecommandlockouts

\usepackage{cite}
\usepackage{amsmath,amssymb,amsfonts}
\usepackage{algorithmic}
\usepackage{graphicx}
\usepackage{textcomp}
\usepackage{xcolor}
\usepackage{acronym}
\def\BibTeX{{\rm B\kern-.05em{\sc i\kern-.025em b}\kern-.08em
    T\kern-.1667em\lower.7ex\hbox{E}\kern-.125emX}}
\begin{document}

\acrodef{5g}[5G]{the fifth generation of mobile technologies}
\acrodef{6g}[6G]{the sixth generation of mobile technologies}
\acrodef{ar}[AR]{Augmented Reality}
\acrodef{ai}[AI]{Artificial Intelligence}
\acrodef{iot}[IoT]{Internet of Things}
\acrodef{ml}[ML]{Machine Learning}
\acrodef{mec}[MEC]{Multi-access Edge Computing}
\acrodef{nfv}[NFV]{Network Function Virtualization}
\acrodef{ovs}[OVS]{Open vSwitch}
\acrodef{qos}[QoS]{Quality of Service}
\acrodef{sdn}[SDN]{Software-Defined Networking}
\acrodef{trl}[TRL]{Technology Readiness Level}
\acrodef{vr}[VR]{Virtual Reality}


\title{Exploring Enabling Technologies for Programmable and Software-Defined Networks: Challenges and Opportunities\\
}
\title{Enabling Technologies for Programmable and Software-Defined Networks: Challenges, Opportunities, and the Path Towards 6G}
\title{Enabling Technologies for Programmable and Software-Defined Networks: Bolstering the Path Towards 6G}


\author{\IEEEauthorblockN{David Carrascal}
\IEEEauthorblockA{\textit{Departamento de Automática} \\
\textit{Universidad de Alcalá}\\
Alcala de Henares (Madrid), Spain \\
david.carrascal@uah.es}
\and
\IEEEauthorblockN{Elisa Rojas}
\IEEEauthorblockA{\textit{Departamento de Automática} \\
\textit{Universidad de Alcalá}\\
Alcala de Henares (Madrid), Spain \\
elisa.rojas@uah.es}
\and
\IEEEauthorblockN{Diego Lopez-Pajares}
\IEEEauthorblockA{\textit{Departamento de Automática} \\
\textit{Universidad de Alcalá}\\
Alcala de Henares (Madrid), Spain \\
diego.lopezp@uah.es}

}
\maketitle


\begin{abstract}

Although the complete scope of the sixth generation of mobile technologies (6G) is still unclear, the prominence of the Internet of Things (IoT) and Artificial Intelligence (AI) / Machine Learning (ML) in the networking field is undeniable. In this regard, key technology enablers for the previous generation, 5G, such as software-defined networking and  network function virtualization, fall short to accomplish the stringent requirements envisioned for 6G verticals. This PhD thesis goes back to basics, by exploring missing functionality gaps in relation to these technologies, in order to provide the ``glue'' for holistic and fully-fledged networking solutions for 6G, aligned with standards and industry recommendations. 
Although ambitious, and in a very early stage, this PhD thesis illustrates an initial design for in-band control in Software-Defined Networking (SDN) that could facilitate the interoperability among constrained IoT devices. 
The current design demonstrates promising results in terms of resource-usage and robustness, which are pivotal features for constrained networks. 
Next steps include the integration of the approach with a real testbed comprised of constrained IoT devices and the implementation of a federated learning environment at the edge. 
\end{abstract}


\section{Introduction} 
\label{sec:intro}

Since the successful launch of the first commercial deployments of \ac{5g}, there is growing interest in the research and innovation results and applications that can be achieved with both \ac{5g} and beyond (i.e. \ac{6g})~\cite{lu20206g,akyildiz20206g}. 
One cornerstone of \ac{5g} networks is founded on moving part of the network intelligence from the core of the network to the edge of the network (users and/or things), enabling what is known as mist/fog/edge computing, or edge-to-cloud continuum~\cite{Milojicic20}, and providing a multitude of advantages and fostering novel applications. 
It is expected that \ac{6g} networks will further build on this paradigm, thanks to the increase in internet-enabled devices~\cite{akyildiz20206g}, 
and at the same time, this paradigm is strongly supported by \ac{sdn} and \ac{nfv} as key technologies in its infrastructure \cite{arch1}~\cite{arch2}. 
Both \ac{sdn} and \ac{nfv} technologies are therefore radically transforming conventional communications networks, as well as posing new challenges and strategies in the upcoming \ac{6g} commercial communications networks~\cite{hu2014survey}. 
However, achieving this new computational model and communication infrastructure still involves overcoming several challenges associated with the foundational components of both the data plane and the control plane of \ac{sdn}/\ac{nfv}-based communication networks. 
For instance, few research efforts consider the potential constraints of memory and energy of \ac{iot} devices for their integration into these softwarized environments. 
Consequently, there is a niche for improvement and optimisation in these technologies, particularly for distributed and constrained environments as those envisioned by the edge-to-cloud continuum. 

This PhD thesis investigates into fundamental technologies for the implementation of \ac{6g} verticals, with a particular emphasis on constrained \ac{iot} devices at the edge and, thus, functional features such as scalability, robustness, and flexibility, are pivotal in the design process. 
In the next sections, we describe the state of the art, objectives and research methodology, some preliminary research ideas, and finally we conclude the paper by providing insights and next steps. 

\section{State of the art} 
\label{sec:sota}

\begin{figure*}[ht!]
    \centering
    \includegraphics[width=0.8\textwidth]{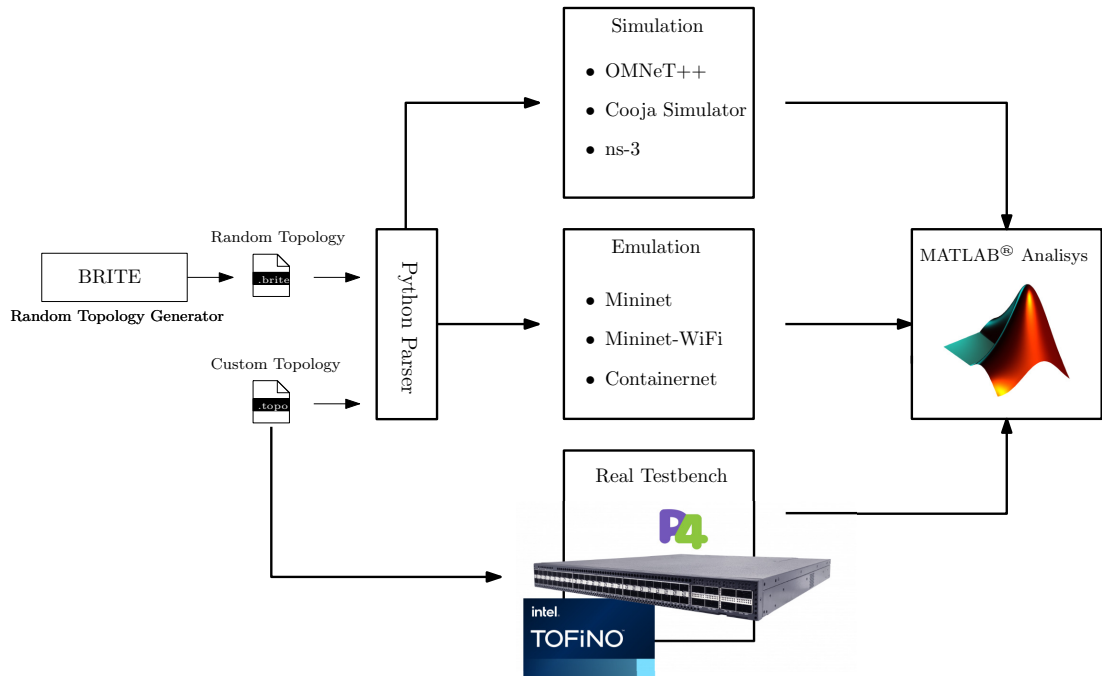}
    \caption{Test architecture design pipeline for the research methodology}
    \label{fig:testPipeline}
\end{figure*}

As a further leap from \ac{5g}, \ac{6g}~\cite{Katz2019} is shaping up to be a revolution in mobile connectivity, with even faster speed, lower latency and greater data capacity. 
But it is not just about speed and capacity: \ac{6g} is also focused on creating smart, sustainable networks to facilitate the green transition, which means a major impact on several key sectors of the economy. 
\ac{6g} verticals include energy, transportation, industry, healthcare, cities and utilities, media and entertainment, tourism and culture~\cite{giordani2020toward}. 
Each of these verticals have specific needs and demands in terms of connectivity and sustainability, and \ac{6g} promises to address these needs and move towards a smarter, more connected society.
For instance, one use case envisions the creation of smart grids for the green transition, which implies greater energy efficiency and the reduction of greenhouse gas emissions in the energy sector~\cite{Uusitalo2021}. 
Moreover, \ac{6g} could improve connectivity in the transportation sector, enabling autonomous vehicles and improved road safety. 
It could also have a significant impact on healthcare, with the creation of more advanced telemedicine networks and improved remote care. 
In the media and entertainment sector, \ac{6g} could improve the quality of video and audio transmission, as well as enable the creation of new augmented and virtual reality experiences. 

In the context of \ac{6g}, \ac{sdn} is a key enabler for the creation of smart and sustainable networks, as it enhances management of network resources (e.g., implementation of green transition policies in real time~\cite{zhang20196g}),  
and it adjusts to the needs of different \ac{6g} verticals, facilitating the design of personalized services and improving the user experience. 

Nevertheless, although \ac{sdn} is already a mature technology, it still suffers from diverse limitations, including security aspects~\cite{Deb22}. 
For example, \ac{6g} envisions multiple type of connectivity links for diverse types of \ac{iot} devices~\cite{Qi2020,Qadir2022}, even working together and located in different areas (earth, heaven and sea)~\cite{Pei23}, but \ac{iot} devices at the edge are usually constrained in memory and energy. 
As such, they cannot support a standard out-of-band \ac{sdn} channel; while using in-band control could be recommended, its deployment is still manually performed in most scenarios since automatization has been barely addressed in literature~\cite{carrascal2023comprehensive}.
Associated verticals that would benefit from scalable and automatic in-band control deployment include sparsely distributed federated learning environments for smart agriculture or local redistribution of computing capabilities for mobile tactical edge environments~\cite{Rojas23}, for example. 
In summary, among other improvements of the \ac{sdn} paradigm, the in-band paradigm is a promising approach for \ac{iot} network management in \ac{6g} environments, and its proper implementation could enable more efficient and scalable management of \ac{iot} devices in the aforementioned different verticals and use cases~\cite{Guo2021,Nguyen2022}. 

\section{Objectives \&  research methodology} 
\label{sec:obj}

According to the previous analysis of the state of the art, which is outlined as a mere starting point, the following research questions are contemplated: 
\begin{itemize}
    \item What enabling technologies are missing in the \ac{sdn}/\ac{nfv} puzzle to be applicable to all type of heterogeneous networks?
    \item How to integrate \ac{ai}/\ac{ml} into \ac{sdn} to enhance network behavior and, vice versa, how to leverage \ac{sdn} for advanced \ac{ai}/\ac{ml} services?
    \item How can security and privacy be ensured in a \ac{6g} network composed of \ac{iot} devices?
    \item Is it possible to involve, and share, the resources of end-user devices (with the privacy issues that this raises) such as mobile, \ac{ar} and \ac{vr} devices in the network edge?
    \item To what extent novel structures generated at the data plane could increase network programmability and bolster \ac{sdn}/\ac{nfv}?
\end{itemize}



The research methodology has been adapted and modified according to the specific needs of the research to be carried out. 
This methodology, depicted in Fig.~\ref{fig:testPipeline}, is mainly based on the wide experience of the research group 
and can be divided into the following steps. 
Firstly, a theoretical study of the state of the art and found solutions, analysing their feasibility and performance, in order to take advantage as much as possible of existing works. Our intention is to prove which works are long-term and foundational ideas, to build upon them, and which are not, to look for potential conceptual gaps. 
This thesis is located at a very low \ac{trl} and this initial step is essential. 
Afterwards, we design and implement our ideas in specific test platforms. 
We like to acknowledge results both via simulation and emulation. 
Finally, a test pipeline will be developed to stress the implemented solutions in order to evaluate their behavior under real conditions, their feasibility, scalability and performance. 
Figure~\ref{fig:testPipeline} summarizes this process, listing some of the most common platforms we use and how relying on a random topology generator is pivotal for comprehensive testbenching in early research stages. 


\section{Preliminary research ideas} 
\label{sec:obj}

The very first research ideas of the author's PhD thesis are focused to design and scalable and resilient in-band \ac{sdn} control protocol for constrained mobile devices at the network edge. 
After a comprehensive survey of the state of the art~\cite{carrascal2023comprehensive} and an initial prototype for in-band in wired/wireless hybrid \ac{sdn} scenarios implemented by the PhD student's research group~\cite{Alvarez2022}, the objective is to re-design the approach reducing memory and energy usage as much as possible, while building robust paths and keeping in-band functionality too. 

\begin{figure}[ht!]
    \centering
    \includegraphics[width=\columnwidth]{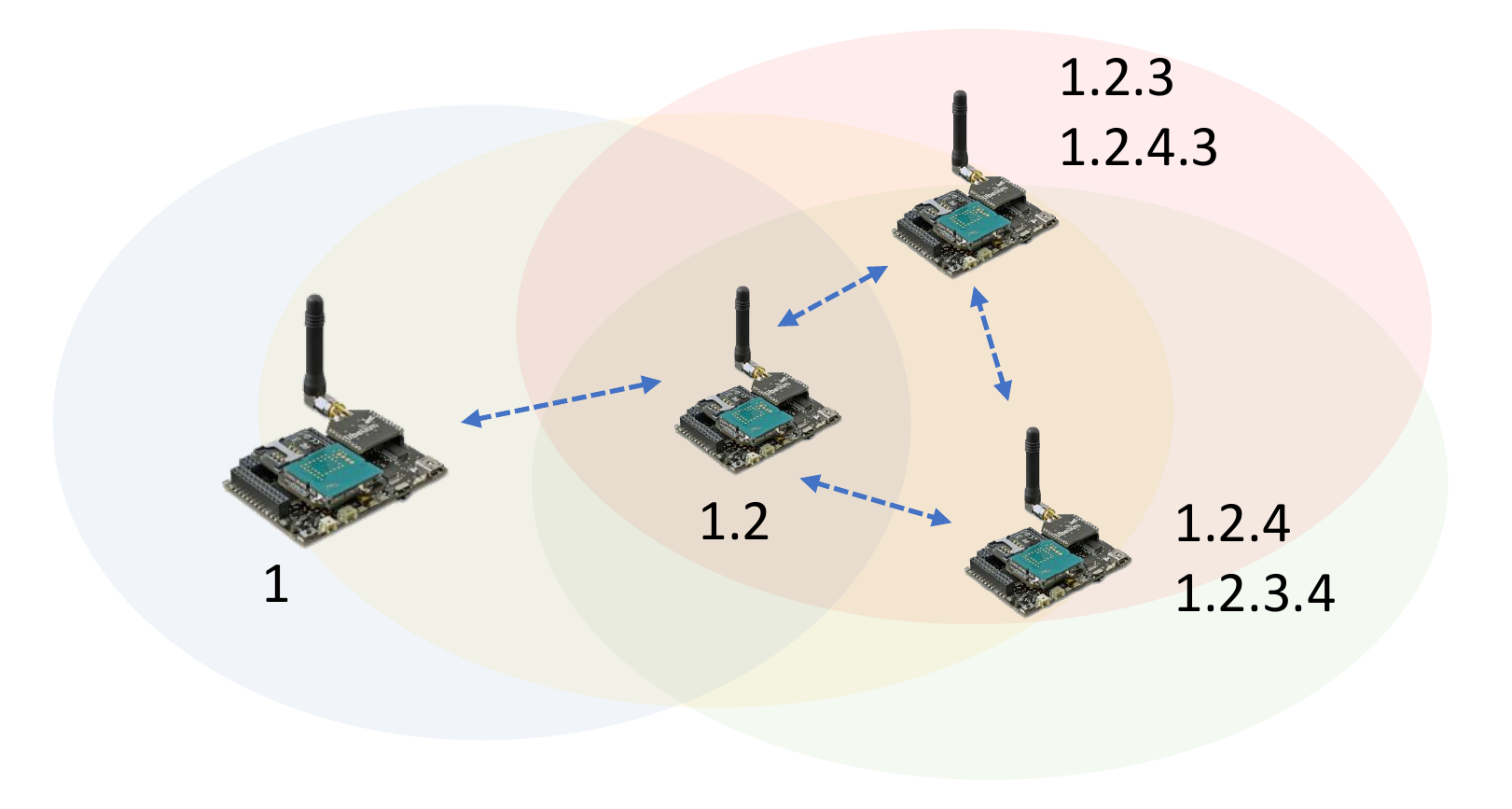}
    \caption{Hierarchical labelling for in-band control in \ac{iot} networks}
    \label{fig:label}
\end{figure}

Since the PhD student had previous knowledge about programmability of \ac{iot} data planes~\cite{carrascal2020analysis} and about \ac{iot} routing protocols~\cite{rojas2021outperforming}, merging both approaches had led to the design of the labelling protocol depicted in Fig.~\ref{fig:label}. 
The idea is as follows: Let us consider a set of \ac{iot} devices, as in Fig.~\ref{fig:label}, one (or more) will behave as gateways (that is, they will be directly connected, or have an already existing path, to an \ac{sdn} controller). 
At the same time, all nodes might broadcast a Hello-like message, so that neighbors are locally aware of what other \ac{iot} devices can be reached. 
With that list of neighbors, gateways will self-determine themselves as such and emit a labelling probe frame to each neighbor, which is received and re-sent with a label update to other neighbors, hence, exploring the whole network. 

In the case of Fig.~\ref{fig:label}, the gateway has label \texttt{1}, and sends label \texttt{1.2} to its only neighbor, by simplifying appending the neighbor ID (2), which is known from the neighbor list, to its own label. 
Afterwards, the node in the middle forwards labels \texttt{1.2.3} and \texttt{1.2.4}, and also a label \texttt{1.2.X} back to the gateway. 
However, the gateway discards such a label following the criterion that its own one \texttt{1} is comprised, as a prefix, in \texttt{1.2.X} (independently of the value of \texttt{X}). 
This latter mechanism grants no forwarding loops are created in the network. 
Finally, nodes \texttt{1.2.3} and \texttt{1.2.4} will also send labels with their neighbors, including \texttt{1.2}, which will discard any label from them to avoid loops, once again. 
More specifically, \texttt{1.2.3} receives \texttt{1.2.4.3}, and \texttt{1.2.4} obtains \texttt{1.2.3.4}. 
The labelling process finalizes when no additional message is exchanged, after the remaining ones are discarded. 
Additionally, the mechanism is compatible with more than one network gateway, even if they assign themselves the same initial label, though this slightly reduces the total amount of assigned labels. 

Once labels are set, the in-band routing is based on the principles of the revision of the IEEE standard 802-2001 published in 2014~\cite{ieee802-2014}, that is, nodes can leverage those labels as their own pseudo-MAC addresses and source-route frames until reaching the gateway. 
For example, node \texttt{1.2.3}/\texttt{1.2.4.3} has two possible routes towards the gateway, one for \texttt{1.2.3}, traversing three nodes, and another for \texttt{1.2.4.3}, through four. 
Selecting one path or another could be based on diverse metrics: fewer hops, smallest latency, etc. 

The main benefits of this approach are: 
\begin{enumerate}
    \item Nodes have as many routes to the gateway as assigned labels (which grants smooth recovery and almost instant route modifications in case of link failure or mobility).
    \item Table size does not depend on the network size, but only on the amount of neighbors and amount of assigned labels (for instance, four entries for node \texttt{1.2.3}, as it has two neighbors and two labels, independently of the rest of the network).
    \item The whole label assignment process simply requires two messages per node: one Hello message and another message to assign labels to neighbors. Of course, in case of high mobility, the number of messages should be tuned. 
\end{enumerate}
Additionally, simulations of the labelling process with Contiki-ng and Cooja demonstrate very low converge times for the whole assignment procedure (of around 1 second for 200-node networks~\cite{rojas2021outperforming}). 

On the other hand, this idea also has limitations about the label length (limited to the length of the MAC address field if the frame is not modified) and potential security issues (concerning the use of Hellos and probe frames). 

To acknowledge the feasibility of the whole in-band protocol, the PhD student is currently working on implementing the design in the specific environment of \ac{sdn}, using the BOFUSS software switch~\cite{Fernandes20}, as it easier to modify than \ac{ovs}. 
The BOFUSS switch is currently being conceptually tested and ported to the latest versions of Linux, to be integrated with Mininet-WiFi. 

Nevertheless, this is just a departing point. 
Afterwards, diverse ideas, explored in parallel, are envisioned as future steps, including: enhancing the security of the in-band protocol, leveraging the protocol for sharing of computational resources in a distributed manner (for example, to implement federated learning approaches in \ac{ar}/\ac{vr} environments), or integrating the idea into a more holistic project (for instance, in the field of ETSI standards for 5G and beyond: \ac{nfv}, \ac{mec}, and the current TeraFlowSDN controller~\cite{Gifre22}). 

\section{Conclusion and next steps}
\label{sec:cf}
This early-stage PhD thesis investigates key technology enablers to bolster the verticals envisioned by \ac{6g} and, more specifically, in fulfilling the missing functionality gaps for \ac{sdn} and programmable networks. 
So far, we have presented the design of a scalable and resilient in-band control approach for \ac{iot}, which will be implemented in a Raspberry Pi-based environment based on BOFUSS. 
Preliminary tests show that the algorithm is capable of creating multiple dynamic paths per node towards the gateway in around 1 second for 200-node networks, which facilitates smooth updates for mobility and link-failure scenarios, and requires small tables (independently of network size) and few exchanged control messages, which is particularly desirable for constrained \ac{iot} environments. 

Regarding future research steps, we will maintain the focus on novel communication protocols and algorithms that enable the programmability of the control and data planes. 
Additionally, the automation of network management through artificial intelligence could be explored further. 
In the long term, as a final step of the thesis, we envisage a prototype of federated vision learning, integrated with \ac{vr} and digital twinning, for a smart agriculture environment. 
We are currently discussing potential applications with two stakeholders from Guadalajara and Madrid, for traditional and innovative farming, respectively. 

\section*{Acknowledgment}

This work was funded by grants from Comunidad de Madrid through projects TAPIR-CM (S2018/TCS-4496) and MistLETOE-CM (CM/JIN/2021-006), by project ONENESS (PID2020-116361RA-I00) of the Spanish Ministry of Science and Innovation, and by Universidad de Alcalá through ``Contratos Predoctorales de Formación de Personal Investigador - FPI-UAH 2022''.

\bibliographystyle{IEEEtran}
\bibliography{refs}

\begin{thebibliography}{10}
\providecommand{\url}[1]{#1}
\csname url@samestyle\endcsname
\providecommand{\newblock}{\relax}
\providecommand{\bibinfo}[2]{#2}
\providecommand{\BIBentrySTDinterwordspacing}{\spaceskip=0pt\relax}
\providecommand{\BIBentryALTinterwordstretchfactor}{4}
\providecommand{\BIBentryALTinterwordspacing}{\spaceskip=\fontdimen2\font plus
\BIBentryALTinterwordstretchfactor\fontdimen3\font minus
  \fontdimen4\font\relax}
\providecommand{\BIBforeignlanguage}[2]{{%
\expandafter\ifx\csname l@#1\endcsname\relax
\typeout{** WARNING: IEEEtran.bst: No hyphenation pattern has been}%
\typeout{** loaded for the language `#1'. Using the pattern for}%
\typeout{** the default language instead.}%
\else
\language=\csname l@#1\endcsname
\fi
#2}}
\providecommand{\BIBdecl}{\relax}
\BIBdecl

\bibitem{lu20206g}
Y.~Lu and X.~Zheng, ``{6G: A survey on technologies, scenarios, challenges, and
  the related issues},'' \emph{Journal of Industrial Information Integration},
  vol.~19, p. 100158, 2020.

\bibitem{akyildiz20206g}
I.~F. Akyildiz, A.~Kak, and S.~Nie, ``{6G and beyond: The future of wireless
  communications systems},'' \emph{IEEE access}, vol.~8, pp.
  133\,995--134\,030, 2020.

\bibitem{Milojicic20}
D.~Milojicic, ``{The Edge-to-Cloud Continuum},'' \emph{Computer}, vol.~53,
  no.~11, pp. 16--25, nov 2020.

\bibitem{arch1}
T.~G. A.~W. Group, ``{The 6G Architecture Landscape European perspective },''
  12 2022.

\bibitem{arch2}
T.~G. A. W. G.~. Hexa-X, ``{Targets and requirements for 6G - initial E2E
  architecture},'' 2022.

\bibitem{hu2014survey}
F.~Hu, Q.~Hao, and K.~Bao, ``{A survey on software-defined network and
  openflow: From concept to implementation},'' \emph{IEEE Communications
  Surveys \& Tutorials}, vol.~16, no.~4, pp. 2181--2206, 2014.

\bibitem{Katz2019}
M.~Katz, M.~Matinmikko-Blue, and M.~Latva-Aho, ``{6Genesis Flagship Program:
  Building the Bridges Towards 6G-Enabled Wireless Smart Society and
  Ecosystem},'' \emph{Proceedings - 2018 10th IEEE Latin-American Conference on
  Communications, LATINCOM 2018}, 1 2019.

\bibitem{giordani2020toward}
M.~Giordani, M.~Polese, M.~Mezzavilla, S.~Rangan, and M.~Zorzi, ``{Toward 6G
  networks: Use cases and technologies},'' \emph{IEEE Communications Magazine},
  vol.~58, no.~3, pp. 55--61, 2020.

\bibitem{Uusitalo2021}
M.~A. Uusitalo, M.~Ericson, B.~Richerzhagen, E.~U. Soykan, P.~Rugeland,
  G.~Fettweis, D.~Sabella, G.~Wikstrom, M.~Boldi, M.~H. Hamon, H.~D. Schotten,
  V.~Ziegler, E.~C. Strinati, M.~Latva-Aho, P.~Serrano, Y.~Zou, G.~Carrozzo,
  J.~Martrat, G.~Stea, P.~Demestichas, A.~Parssinen, and T.~Svensson,
  ``{Hexa-X: The european 6G flagship project},'' \emph{2021 Joint European
  Conference on Networks and Communications and 6G Summit, EuCNC/6G Summit
  2021}, pp. 580--585, 6 2021.

\bibitem{zhang20196g}
Z.~Zhang, Y.~Xiao, Z.~Ma, M.~Xiao, Z.~Ding, X.~Lei, G.~K. Karagiannidis, and
  P.~Fan, ``{6G wireless networks: Vision, requirements, architecture, and key
  technologies},'' \emph{IEEE Vehicular Technology Magazine}, vol.~14, no.~3,
  pp. 28--41, 2019.

\bibitem{Deb22}
\BIBentryALTinterwordspacing
R.~Deb and S.~Roy, ``{A comprehensive survey of vulnerability and information
  security in SDN},'' \emph{Computer Networks}, vol. 206, p. 108802, 2022.
  [Online]. Available:
  \url{https://www.sciencedirect.com/science/article/pii/S1389128622000299}
\BIBentrySTDinterwordspacing

\bibitem{Qi2020}
Q.~Qi, X.~Chen, C.~Zhong, and Z.~Zhang, ``{Integration of energy, computation
  and communication in 6G cellular internet of things},'' \emph{IEEE
  Communications Letters}, vol.~24, pp. 1333--1337, 6 2020.

\bibitem{Qadir2022}
Z.~Qadir, K.~N. Le, N.~Saeed, and H.~S. Munawar, ``{Towards 6G Internet of
  Things: Recent advances, use cases, and open challenges},'' \emph{ICT
  Express}, 6 2022.

\bibitem{Pei23}
J.~Pei, S.~Li, Z.~Yu, L.~Ho, W.~Liu, and L.~Wang, ``{Federated Learning
  Encounters 6G Wireless Communication in the Scenario of Internet of
  Things},'' \emph{IEEE Communications Standards Magazine}, vol.~7, no.~1, pp.
  94--100, 2023.

\bibitem{carrascal2023comprehensive}
D.~Carrascal, E.~Rojas, J.~M. Arco, D.~Lopez-Pajares, J.~Alvarez-Horcajo, and
  J.~A. Carral, ``{A Comprehensive Survey of In-Band Control in SDN: Challenges
  and Opportunities},'' \emph{Electronics}, vol.~12, no.~6, p. 1265, 2023.

\bibitem{Rojas23}
E.~Rojas, D.~Lopez-Pajares, J.~Alvarez-Horcajo, and S.~Llopis~S{\'a}nchez,
  ``The cloud continuum for military deployable networks: Challenges
  and opportunities,'' in \emph{Computer Security. ESORICS 2022 International
  Workshops}.\hskip 1em plus 0.5em minus 0.4em\relax Cham: Springer
  International Publishing, 2023, pp. 500--519.

\bibitem{Guo2021}
F.~Guo, F.~R. Yu, H.~Zhang, X.~Li, H.~Ji, and V.~C. Leung, ``{Enabling Massive
  IoT Toward 6G: A Comprehensive Survey},'' \emph{IEEE Internet of Things
  Journal}, vol.~8, pp. 11\,891--11\,915, 8 2021.

\bibitem{Nguyen2022}
D.~C. Nguyen, M.~Ding, P.~N. Pathirana, A.~Seneviratne, J.~Li, D.~Niyato,
  O.~Dobre, and H.~V. Poor, ``{6G Internet of Things: A Comprehensive
  Survey},'' \emph{IEEE Internet of Things Journal}, vol.~9, pp. 359--383, 1
  2022.

\bibitem{Alvarez2022}
\BIBentryALTinterwordspacing
J.~Alvarez-Horcajo, I.~Martinez-Yelmo, E.~Rojas, J.~A. Carral, and
  D.~Carrascal, ``{ieHDDP: An Integrated Solution for Topology Discovery and
  Automatic In-Band Control Channel Establishment for Hybrid SDN
  Environments},'' \emph{Symmetry}, vol.~14, no.~4, p. 756, Apr 2022. [Online].
  Available: \url{http://dx.doi.org/10.3390/sym14040756}
\BIBentrySTDinterwordspacing

\bibitem{carrascal2020analysis}
D.~Carrascal, E.~Rojas, J.~Alvarez-Horcajo, D.~Lopez-Pajares, and
  I.~Mart{\'\i}nez-Yelmo, ``{Analysis of P4 and XDP for IoT Programmability in
  6G and Beyond},'' \emph{IoT}, vol.~1, no.~2, pp. 605--622, 2020.

\bibitem{rojas2021outperforming}
E.~Rojas, H.~Hosseini, C.~Gomez, D.~Carrascal, and J.~R. Cotrim,
  ``{Outperforming RPL with scalable routing based on meaningful MAC
  addressing},'' \emph{Ad Hoc Networks}, vol. 114, p. 102433, 2021.

\bibitem{ieee802-2014}
``{IEEE Standard for Local and Metropolitan Area Networks: Overview and
  Architecture},'' \emph{IEEE Std 802-2014 (Revision to IEEE Std 802-2001)},
  pp. 1--74, 2014.

\bibitem{Fernandes20}
\BIBentryALTinterwordspacing
E.~L. Fernandes, E.~Rojas, J.~Alvarez-Horcajo, Z.~L. Kis, D.~Sanvito,
  N.~Bonelli, C.~Cascone, and C.~E. Rothenberg, ``{The road to BOFUSS: The
  basic OpenFlow userspace software switch},'' \emph{Journal of Network and
  Computer Applications}, vol. 165, p. 102685, 2020. [Online]. Available:
  \url{https://www.sciencedirect.com/science/article/pii/S1084804520301594}
\BIBentrySTDinterwordspacing

\bibitem{Gifre22}
L.~Gifre, D.~King, A.~Farrel, R.~Casellas, R.~Martínez, J.-P.
  Fernández-Palacios, O.~González-de Dios, J.-J. Pedreno-Manresa,
  A.~Autenrieth, R.~Muñoz, and R.~Vilalta, ``{Experimental Demonstration of
  Transport Network Slicing with SLA Using the TeraFlowSDN Controller},'' in
  \emph{2022 European Conference on Optical Communication (ECOC)}, 2022, pp.
  1--4.

\end{thebibliography}

\end{document}